\documentclass{elsart3}

\usepackage{graphicx}
\usepackage{amsmath}

\usepackage{amssymb}

\def\etal{{\it et\ al.\ }}

\newcommand{\lsim}
 {\ \raise.35ex\hbox{$<$}\kern-0.75em\lower.5ex\hbox{$\sim$}\ }
\newcommand{\gsim}
 {\ \raise.35ex\hbox{$>$}\kern-0.75em\lower.5ex\hbox{$\sim$}\ }
\def\journal #1#2#3#4{#1 {\bf #2}, #3 (#4)}
\def\PR{Phys.\ Rev.}
\def\PRB{Phys.\ Rev.\ B}
\def\PRL{Phys.\ Rev.\ Lett.}

\def\JPSJ{J.\ Phys.\ Soc.\ Jpn.}

\def\PTP{Prog.\ Theor.\ Phys.}

\def\EPL{Europhys.\ Lett.}

\begin{document}

\begin{frontmatter}



\title{Doublon-Holon Binding Effects on Mott Transitions 
in Two-Dimensional Bose Hubbard Model}


\author[label1]{Hisatoshi Yokoyama\corauthref{cor1}}
\author[label2]{Masao Ogata}

\corauth[cor1]{Corresponding author. TEL./fax: +81-22-795-6444 \\
E-mail: yoko@cmpt.phys.tohoku.ac.jp (H.~Yokoyama)}

\address[label1]{Department of Physics, Tohoku University, Sendai 980-8578, Japan.}
\address[label2]{Department of Physics, University of Tokyo, Bunkyo-ku, Tokyo 113-0033, Japan.}

\begin{abstract}

A mechanism of Mott transitions in a Bose Hubbard model on a square lattice 
is studied, using a variational Monte Carlo method. 
Besides an onsite correlation factor, we introduce a four-body doublon-holon 
factor into the trial state, which considerably improves the variational 
energy and can appropriately describe a superfluid-insulator transition. 
Its essense consists in binding (and unbinding) of a doublon to a holon 
in a finite short range, identical with the cases of fermions. 
The features of this transition are qualitatively different from those 
of Brinkman-Rice-type transitions. 
\end{abstract}

\end{frontmatter}

\null
\pagebreak[4]

{\it 1.\ Introduction}:\ 
After early theoretical studies of Mott or superfluid-insulator 
transitions in interacting Bose systems \cite{Fisher}, 
an experimental example has been realized using an ultracold 
dilute gas of bosonic atoms in an optical lattice \cite{Greiner}. 
The essence of this system is considered to be captured \cite{Jaksch} 
by a Bose Hubbard model. 
This basic model has been studied with various methods; for square 
lattices, properties of $T_{\rm c}$, superfluid density, etc.~were 
studied, applying a quantum Monte Carlo method to small systems 
(mainly $6\times6$ square lattice) \cite{Trivedi}, and a ground-state 
phase diagram in a plane of chemical potential and interaction 
strength was obtained, using a strong-coupling expansion \cite{Monien}. 
These studies estimated the critical interaction strength of 
Mott transitions at $U_{\rm c}/t=16.4\pm 0.8$ and 16.69, respectively, 
for the particle density of $n=1$ ($n=N_{\rm e}/N$ with $N_{\rm e}$: 
particle number, $N$: site number) at $T=0$. 
Thus, the existence of a Mott transition has been embodied, but the 
mechanism of the transition is still not clear. 
\par

Variational Monte Carlo approaches are very useful to understand 
the mechanism of the Mott transition, because one can directly and 
exactly treat wave functions. 
For the Bose Hubbard model, wave functions with only onsite correlation 
factors, which corresponds to the celebrated Gutzwiller wave function 
(GWF, $\Psi_{\rm G}$) 
for fermions \cite{Gutz}, were studied first \cite{BHM-GWF}. 
In contrast to for fermions, GWF for bosons is solved analytically 
without additional mean-field-type approximations \cite{GA} for 
arbitrary dimensions, and yield a Brinkman-Rice-type (BR) transition 
\cite{BR} at $U=U_{\rm BR}$. 
In BR transitions, all the lattice sites are occupied with exactly one 
particle and the hopping completely ceases in the insulating regime, 
namely, $\Psi_{\rm G}\rightarrow\prod_{j=1}^Nb_j^\dag|0\rangle$ and 
$E=0$ for $U>U_{\rm BR}$. 
This result is caused by an oversimplified setup of the wave function, 
in which the effect of density fluctuation should be included. 
In this work, we introduce a doublon-holon binding correlation factor 
into the trial function, following previous studies for fermions. 
Thereby, we can describe a superfluid-insulator transition more 
appropriately. 
\par


{\it 2.\ Formulation}:\ 
We consider a spinless Bose-Hubbard model on a square lattice, 
\begin{equation}
H = -t \sum_{\langle ij\rangle} {(b_i^\dag b_j + b_j^\dag b_i)} 
+ \frac{U}{2}\sum_j n_j(n_j-1), 
\label{eq:model} 
\end{equation}
where $b^\dag_j$ is a creation operator of a boson at site $j$, 
$n_j=b^\dag_jb_j$ and $t$, $U>0$. 
Here, $\langle ij\rangle$ denotes a nearest-neighbor-site pair; 
the definition of $t$ is a half of what was given in some 
literatures \cite{Trivedi,BHM-GWF}. 
In this work, we restrict to $n\sim 1$, 
because we would like to consider the most simple case of Mott 
transitions. 
The cases of other commensurate densities ($n\ge 2$) must be 
essentially identical. 
\par

We study this model eq.~(\ref{eq:model}) through a variation theory. 
As a trial wave function, we use a Jastrow type of 
$
\Psi_Q=P_Q P_{\rm G}\Phi
$
(QWF), following previous studies for fermions \cite{KHF,YS3}. 
Here, $\Phi$ is the ground state of noninteracting (completely coherent) 
bosons, namely $\Phi=1$, $P_{\rm G}$ is an onsite projector corresponding 
to the famous Gutzwiller factor for fermions 
\cite{Gutz}: 
$ 
P_{\rm G}(g)=g^D, 
$
with
$
D=\sum_i n_{i}(n_{i}-1)/2. 
$ 
As we repeatedly showed \cite{YS3,WataSNS}, intersite correlation factors 
are indispensable for appropriate descriptions of interacting systems. 
In particular near half filling, a four-body doublon-holon correlation 
factor $P_Q$ is crucial to explain the mechanism of Mott transitions 
\cite{YPTP,YTOT,Wata}. 
For $S=1/2$ fermions, $P_{\tilde Q}$ is explicitly written as, 
\begin{equation} 
P_{\tilde Q}(\mu)=(1-\mu)^{\tilde Q}
\label{eq:pqf}
\end{equation}
with $0\le\mu\le 1$ and
\begin{equation}
\tilde Q=\sum_i\prod_{\tau} \left[d_i(1-e_{i+\tau})+e_i(1-d_{i+\tau})\right], 
\label{eq:qf}
\end{equation}
where $d_i$ ($=n_{i\uparrow}n_{i\downarrow}$) and $e_i$ 
[$=(1-n_{i\uparrow})(1-n_{i\downarrow})$] are the doublon and holon 
operators respectively, $i$ runs over all the sites, and $\tau$ the 
four nearest-neighbor sites of the site $i$. 
When the variational parameter $\mu$ vanishes, $\Psi_Q$ is reduced to the 
GWF, $\Psi_{\rm G}=P_{\rm G}\Phi$, in which 
doublons and holons can move independently. 
On the other hand, in the limit of $\mu\rightarrow 1$, a doublon becomes 
bound to a holon in nearest-neighbor sites. 
Consequently, plus (holons) and minus (doublons) charge carriers 
completely cancels, indicating a metal-insulator transition. 
\par

In this work, we extend eq.~(\ref{eq:pqf}) to bosons, and include 
the correlation between diagonal (second-nearest)-neighbor sites: 
\begin{equation}
P_Q(\mu,\mu')=(1-\mu)^Q(1-\mu')^{Q'}, 
\label{eq:pqb}
\end{equation}
where the primes ($'$) denote the cases of diagonal neighbors. 
In eq.~(\ref{eq:pqb}), $Q$ has the same form as $\tilde Q$ in 
eq.~(\ref{eq:qf}), but the doublon operator $d_i$ is replaced by 
a multiplon operator $m_i$, which yields $1$ (0) if the site $i$ is 
multiply occupied (otherwise). 
Near Mott transitions, $P_Q$ of eq.~(\ref{eq:pqb}) is substantially 
a doublon-holon factor, because the probability density of multiplon 
with more than two bosons is almost zero for such large values of $U/t$. 
\par

To estimate expectation values accurately, we use a variational Monte 
Carlo method \cite{McMillan,YS1,Umrigar} of fixed particle numbers. 
First, we optimize the variational parameters, $g$, $\mu$ and $\mu'$, 
simultaneously for each set of $U/t$, $n$ and $L$, and then calculate 
physical quantities with the optimized parameters. 
Through the optimization process, we average substantially several 
million samples, which reduces statistical errors in the total energy 
typically to the order of $10^{-4}t$. 
To check system-size dependence, particularly near phase transitions, 
we employ square lattices of $N=L\times L$ sites up to $N=1024$ 
($L=32$) with the periodic-periodic boundary conditions. 
\par


{\it 3.\ Results}:\ 
\begin{figure}
\begin{center}
\includegraphics[width=7cm,clip]{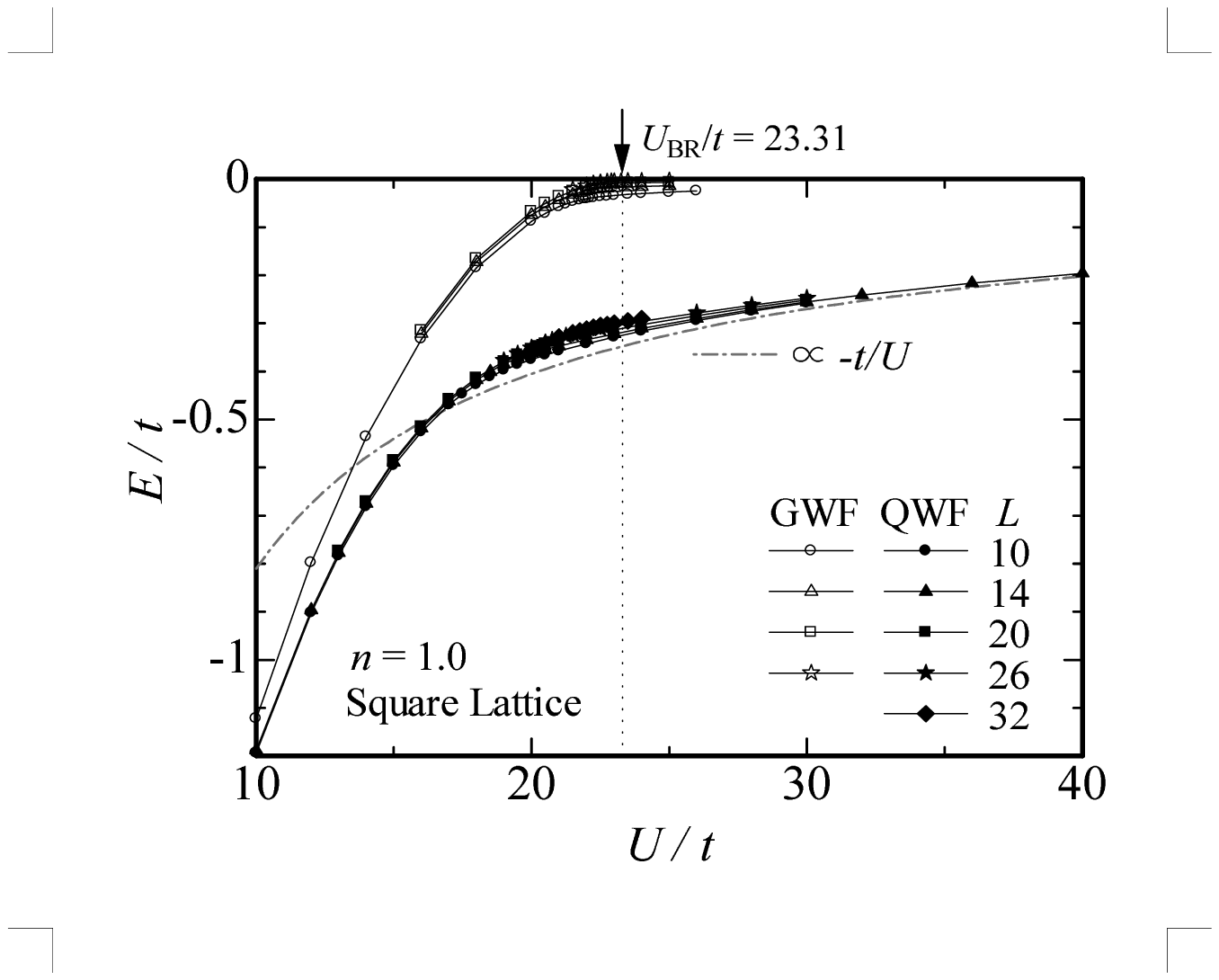}
\end{center}
\vspace{-0.5cm}
\caption{\label{fig:EvsU} 
Comparison of total energy per site between GWF and QWF as a function of 
$U/t$ for several system sizes. 
The particle density is commensurate ($n=1.0$).
For GWF, a Brinkman-Rice-type transition occurs at $U=U_{\rm BR}$. 
The dash-dotted line is a curve proportional to $-t/U$. 
}
\end{figure}
%
We start with comparison of the total energies $E/t$ between GWF and QWF, 
which are shown in Fig.~\ref{fig:EvsU} for $n=1$. 
As mentioned, in GWF \cite{BHM-GWF}, a BR transition 
occurs at $U_{\rm BR}/t=12+8\sqrt{2}=23.31...$; for $U>U_{\rm BR}$, 
each site is occupied by exactly one particle and hopping or density 
fluctuation completely ceases. 
Consequently, $E/t$ vanishes in the insulating regime. 
On the side of Bose fluid ($U<U_{\rm BR}$), $E/t$ vanishes as 
$\propto(1-U/U_{\rm BR})^2$, meaning this transition is a continuous 
type. 
The difference of the two functions is small for small $U/t$ ($\lsim 15$), 
but it becomes conspicuous as $U/t$ approaches Mott critical values 
($U/t\gsim 20$). 
Thus, QWF is a considerably improved function in the point of the 
variation principle. 
\par

\begin{figure}
\begin{center}
\includegraphics[width=7cm,clip]{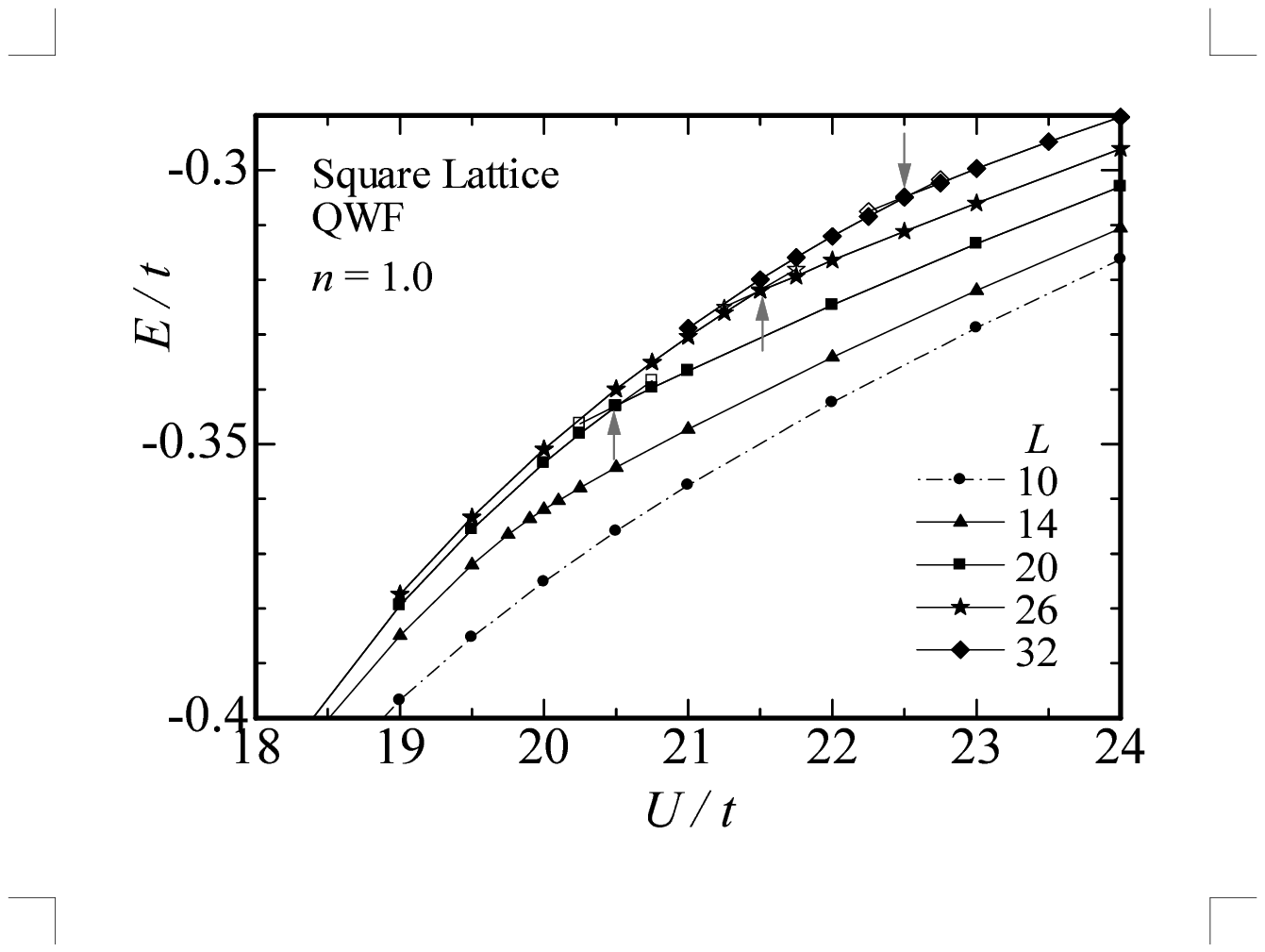}
\end{center}
\vspace{-0.5cm}
\caption{\label{fig:EvsUm} 
Magnification of total energy of QWF near Mott transitions arising 
at $U=U_{\rm c}$ indicated by arrows. 
Data for five system sizes are plotted; for $L\ge 20$, double-minimum 
or cusp behavior is observed at $U=U_{\rm c}$. 
Solid symbols denote the energies of the optimized states, and open
symbols near $U_{\rm c}$ metastable states. 
}
\end{figure}
%
As we will discuss in detail shortly, QWF also exhibits a 
superfluid-insulator transition, but its behavior is qualitatively 
different from that of GWF. 
In contrast to $E$ of GWF, $E$ of QWF in the insulating regime 
($U>U_{\rm c}$) does not vanish but is proportional to $-t^2/U$ 
as seen in Fig.~\ref{fig:EvsU}, which behavior is expected from 
strong-correlation theories, namely, the density fluctuation does 
not cease but is restricted to a finite short range. 
This is the essence of the mechanism of Mott transition owing to the
doublon-holon binding. 
In Fig.~\ref{fig:EvsUm}, we show the magnification of $E/t$ of QWF for 
$U\sim U_{\rm c}$. 
For large $L$ ($>20$), we find, near $U=U_{\rm c}$, double-minimum 
structure of $E/t$ in the space of variational parameters, meaning 
this transition is first order, although such structure cannot be 
confirmed for $L\le 14$ by our VMC calculations. 
\par

\begin{figure}
\begin{center}
\includegraphics[width=7cm,clip]{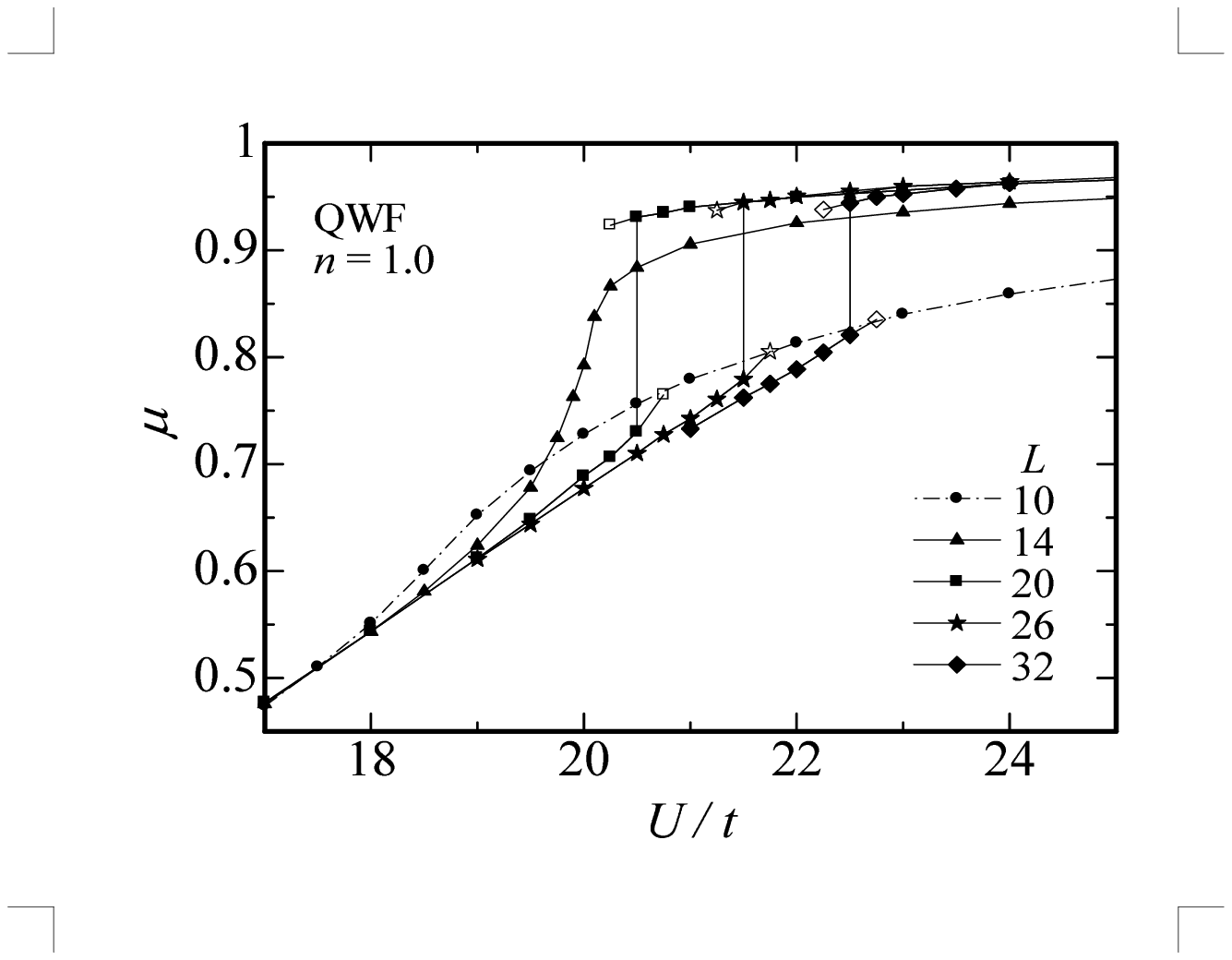}
\end{center}
\vspace{-0.5cm}
\caption{\label{fig:paramum} 
Behavior of doublon-holon binding parameter near $U_{\rm c}$. 
Data for five system sizes are plotted; for $L\ge 20$, discontinuities 
are observed at $U=U_{\rm c}$. 
Solid and open symbols have the same meaning as in Fig.~\ref{fig:EvsUm}. 
}
\end{figure}
%
\begin{figure}
\begin{center}
\includegraphics[width=7cm,clip]{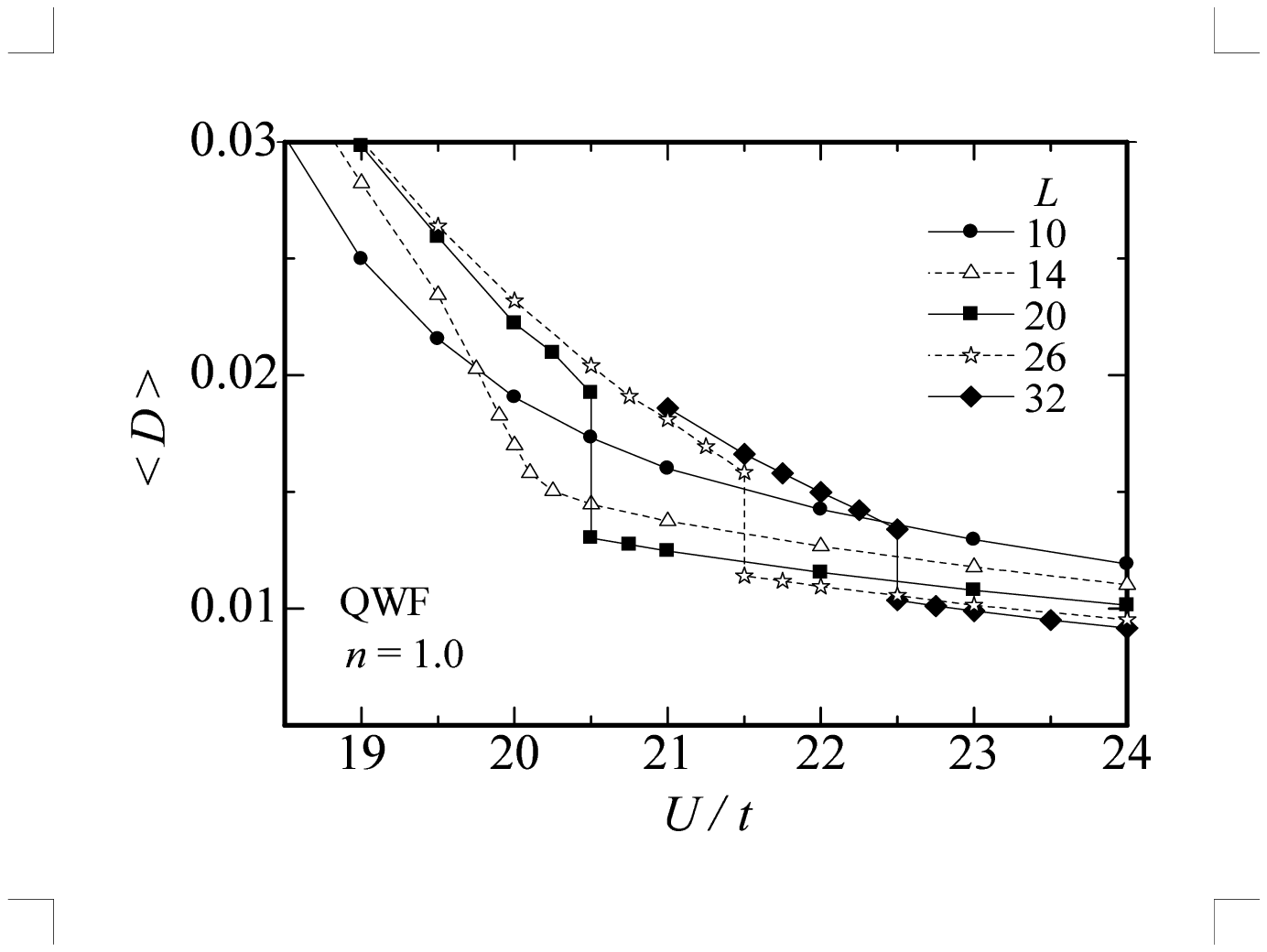}
\end{center}
\vspace{-0.5cm}
\caption{\label{fig:DvsUm} 
Behavior of $\langle D\rangle$, which is substantially the doublon 
density and an order parameter of Mott transitions. 
Data are shown only for the optimized states. 
}
\end{figure}
%
The first-order features are more easily found in the variational 
parameters and in some quantities. 
In Fig.~\ref{fig:paramum}, we plot the optimized doublon-holon binding 
parameter $\mu$ near $U_{\rm c}$. 
For $L\ge 20$, there are clear discontinuities at $U=U_{\rm c}$; the 
large values of $\mu$ for $U>U_{\rm c}$ indicate that a doublon and 
a holon are tightly bound in nearest-neighbor sites. 
In Fig.~\ref{fig:DvsUm}, we show the average of $D$, which is, near 
$U_{\rm c}$, virtually identical with the doublon density, namely, 
an order parameter of Mott transitions. 
The discontinuities of this quantity at $U_{\rm c}$ corroborate 
a first-order Mott transition. 
\par

%
\begin{figure}
\begin{center}
\includegraphics[width=7cm,clip]{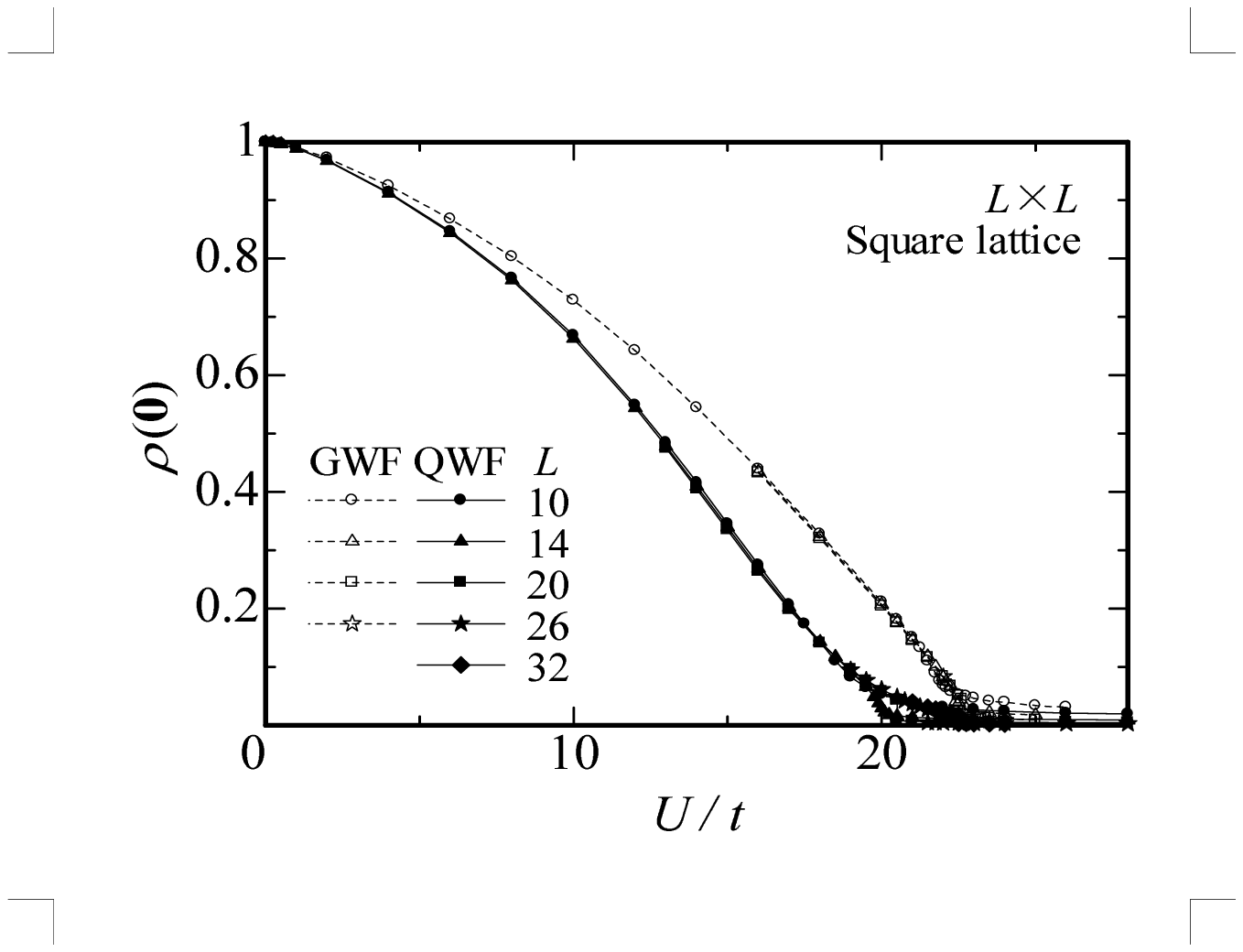}
\end{center}
\vspace{-0.5cm}
\caption{\label{fig:cor0} 
The occupation rate of the ${\bf k}=(0,0)$ level for the two wave 
functions is depicted as a function of $U/t$.  
Data for several system sizes are simultaneously plotted. 
}
\end{figure}
%
\begin{figure}
\begin{center}
\includegraphics[width=7cm,clip]{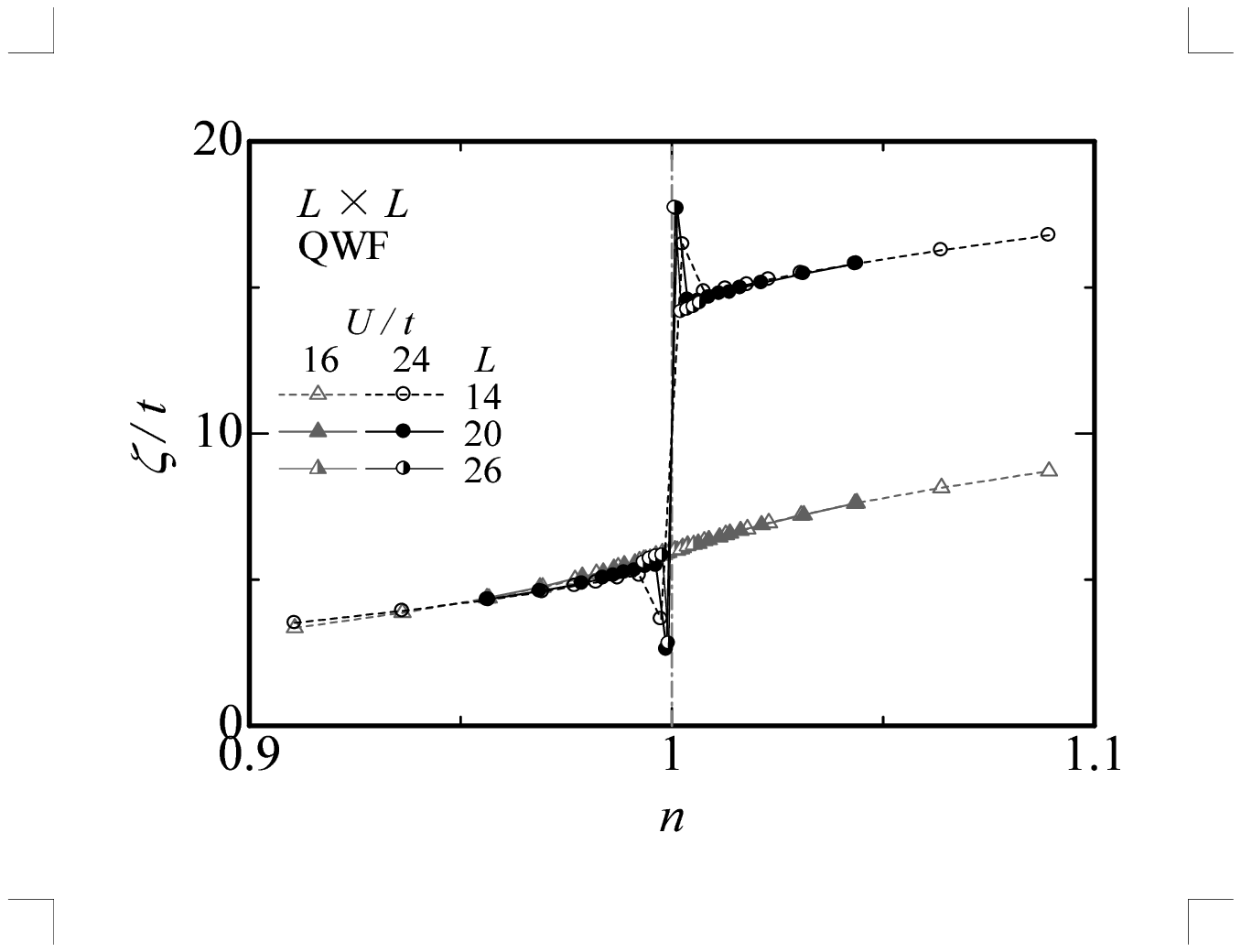}
\end{center}
\vspace{-0.5cm}
\caption{\label{fig:cp} 
The Behavior of chemical potential, $\zeta/t$, as a function of particle 
density near $n=1$ is shown for two values of $U/t$, 16 in the superfluid 
regime and 24 in the insulating regime. 
Data of three system sizes are simultaneously plotted for each $U/t$. 
The singular behavior in $\zeta$ at $n=1$ for $U>U_{\rm c}$ stems from 
the singularly low $E$ at $n=1$ as a function of $n$.  
}
\end{figure}
%
Finally, let us consider a couple of properties of this Mott transition. 
In Fig.~\ref{fig:cor0}, we show the occupation rate, 
$\rho({\bf k})=\langle b_{\bf k}^\dag b_{\bf k}\rangle/N$,
of the lowest-energy level, ${\bf k}={\bf 0}=(0,0)$ versus $U/t$. 
Although $\rho({\bf 0})$ does not directly indicate the superfluid 
density, it must be a good index of superfluidity. 
For the noninteracting case ($U/t=0$), all the particles fall in 
the ${\bf k}={\bf 0}$ level. 
As the interaction becomes strong, $\rho({\bf 0})$ decreases at first 
gradually and drops discontinuously to the order of $1/N$ at $U_{\rm c}$. 
Thus, the superfluidity vanishes at the transition. 
In Fig.~\ref{fig:cor0}, we plot the chemical potential, 
$\zeta=\partial E/\partial n$, estimated from finite differences, 
as a function of $n$. 
In the superfluid regime slightly below $U_{\rm c}$ ($U/t=16$, gray 
symbols), $\zeta$ is a smooth function of $n$ even at $n=1$, indicating 
the state is gapless in density excitation. 
On the other hand in the insulating regime slightly above $U_{\rm c}$
($U/t=24$, black symbols), $\zeta$ has a large discontinuity at $n=1$; 
a density excitation gap opens for $U>U_{\rm c}$ at $n=1$.
The gap behavior is also confirmed by the density correlation 
function $N({\bf q})$ for small $|{\bf q}|$ (not shown).
\par

%

{\it 4.\ Discussions}:\ 
In this proceedings, we have found that a wave function with 
doublon-holon correlation factor, $\Psi_Q$, qualitatively improve 
the description of a Mott transition also in a Bose system. 
Thus, it is probable that the mechanism of Mott transitions for 
bosons is basically identical to that for fermions. 
It is urgent to compare theoretical results with experiments 
particularly of optical lattices. 
We have left many issues to be discussed, which will be published 
elsewhere soon. 
When main calculations here were finished, we became aware that 
a similar wave function had been studied recently \cite{Sorella}. 
\par




\begin{thebibliography}{99}

\bibitem{Fisher} For instance, M.~P.~A.~Fisher \etal, 
\journal{\PRB}{40}{546}{1989}. 

\bibitem{Greiner} M.~Greiner \etal, \journal{Nature}{415}{39}{2002}. 

\bibitem{Jaksch} D.~Jaksch \etal, \journal{\PRB}{81}{3108}{1998}.

\bibitem{Trivedi} W.~Krauth, N.~Trivedi, 
\journal{\EPL}{14}{627}{1991}.

\bibitem{Monien} N.~Elstner and H.~Monien, 
\journal{\PRB}{59}{12184}{1999}. 

\bibitem{Gutz} M.~Gutzwiller, \journal{\PRL}{10}{159}{1963}.

\bibitem{BHM-GWF} W.~Krauth, M.~Caffarel, J.~-P.~Bouchaud, 
\journal{\PRB}{45}{3137}{1992}; 
D.~S.~Rokhsar and B.~G.~Kotliar, 
\journal{\PRB}{44}{10328}{1991}. 

\bibitem{GA} M.~Gutzwiller, \journal{\PR}{137}{A1726}{1965}.

\bibitem{BR} W.~Brinkman, T.~M.~Rice, 
\journal{\PRB}{2}{1324}{1970}.

\bibitem{KHF} T.~A.~Kaplan, P.~Horsch, P.~Fulde, 
\journal{\PRL}{49}{889}{1982}. 

\bibitem{YS3} H.~Yokoyama, H.~Shiba, \journal{\JPSJ}{59}{3669}{1990}. 

\bibitem{WataSNS} T.~Watanabe, H.~Yokoyama, Y.~Tanaka, J.~Inoue, 
to be published in this volume. 

\bibitem{YPTP} H.~Yokoyama, \journal{\PTP}{108}{59}{2002}

\bibitem{YTOT} H.~Yokoyama, \etal, \journal{\JPSJ}{73}{1119}{2004}; 
H.~Yokoyama, M.~Ogata, Y.~Tanaka, \journal{\JPSJ}{75}{114706}{2006}. 

\bibitem{Wata} T.~Watanabe, \etal, \journal{\JPSJ}{75}{074707}{2006}. 

\bibitem{McMillan} W.~L.~McMillan, \journal{\PR}{138}{A442}{1965}. 

\bibitem{YS1} H.~Yokoyama, H.~Shiba, \journal{\JPSJ}{56}{1490}{1987}. 

\bibitem{Umrigar} C.~J.~Umrigar, K.~G.~Wilson, J.~W.~Wilkins, 
\journal{\PRL}{60}{1719}{1988}. 

\bibitem{Sorella} M.~Capello \etal, preprint (cond-mat/0705.2684).

\end{thebibliography}
\end{document}